\documentstyle[epsfig]{aipproc}
\begin{document}
\hskip 11cm SU-4240-713
\title{Light Scalar Mesons} 
\author{Deirdre Black, Amir H. Fariborz and Joseph Schechter}
\address{ Department of Physics, Syracuse University, Syracuse, New
York 13244-1130, USA.}
%\lefthead{LEFT head}
%\righthead{RIGHT head}
\maketitle
\begin{abstract}

We review how a certain effective chiral Lagrangian approach to
$\pi\pi$ scattering, $\pi K$ scattering and $\eta^\prime \rightarrow
\eta\pi\pi$ decay provides evidence for the existence of light scalars
$\sigma(550)$ and $\kappa(900)$ as well as describing the $f_0(980)$
and the $a_0(980)$. An attempt to fit these into a nonet suggests
that their structure is closer to a dual quark--dual antiquark
than to a quark--antiquark. A possible mechanism to explain the next
higher mass scalar nonet is also proposed.

\end{abstract}
\section{Introduction}
The possible existence of light scalar mesons (with masses less than
about 1 GeV) has been a controversial subject for roughly forty years.
There are two aspects: the extraction of the scalar properties from
experiment and their underlying quark substructure. Because the 
$J=0$ channels may contain strong competing contributions, such resonances
may not necessarily dominate their amplitudes and could be hard to
``observe". In such an instance their verification would be
linked to the model used to describe them. The last few
years have seen a revival of interest in this
area. As examples, three models for the underlying quark structure
have been discussed by many authors, including other contributors
to this workshop\cite{workshop}: i) the $K{\bar K}$ molecule model 
\cite{kkbar},
ii) the $q{\bar q}$ model with strong meson-meson interactions
(or "unitarized quark model")\cite{uqm}, iii) the intrinsic $ qq{\bar
q}{\bar q}$
model (Jaffe type\cite{jaffe}). These models have the common feature that
four quarks
are involved in some form; all are different from the ``simple"
$q{\bar q}$ model. Clearly, the elucidation of the structure of unusual
low lying states can be expected to increase our understanding of
non-perturbative QCD.

The present approach is based on comparing with experiment, the 
predictions for $\pi\pi$ scattering, $\pi K$ scattering and $\eta^\prime
\rightarrow \eta\pi\pi$ decay from a phenomenological chiral
Lagrangian containing particles of mass comparable to the energy regime of
interest. These studies seem to require for consistency the existence 
of two isoscalars $\sigma(550)$ and $f_0(980)$, an isospinor $\kappa(900)$
and an isovector $a_0(980)$ with given properties. Note that in the
effective Lagrangian approach, the quark substructure of the scalars is
not specified. In particular a nonet field can {\it a priori}
represent either $q{\bar q}$ or $qq{\bar q}{\bar q}$ (or even more
complicated) states. From this point of view our approach is
``model independent".

Section II contains a brief summary of our scattering model in the
$\pi\pi$ case. The generalization to the $\pi K$ case and to
$\eta^{\prime} \rightarrow \eta\pi\pi$ decay is even more briefly
summarized in section III. Section IV deals with the ``family properties"
of the nonet made up from the scalars we need. The model of
``ideal mixing" for meson nonets is reviewed and generalized to include
``dual ideal mixing". The realistic situation is noted to be closer to
dual rather than ordinary ideal mixing. Finally, in section V, a
possible mechanism is proposed to explain some puzzling features
of a presumably more conventional next--to--lowest--lying scalar nonet.

\section{pi pi scattering}
    The most difficult partial wave amplitude to explain is just
the scalar channel with $I=J=0$. Our notation for the partial wave is
$T^I_J(s) = R^I_J + iI^I_J$. The complicated shape of the experimentally
obtained $R^0_0(s)$ shown in Figs. 2 and 3 below suggests that
resonances are present. Close to threshold, the chiral perturbation theory 
approach, which essentially supplies a Taylor expansion of the amplitude,
is very accurate. However explaining the data shown to about 1.2 GeV
would appear to require a prohibitively high order of expansion in this
scheme. Thus we sacrifice some accuracy near threshold and use instead an
expansion of the invariant amplitude in terms of resonance exchange
diagrams (including contact terms needed for chiral symmetry). This
holds the possibility of achieving a fit to experiment over a larger
energy regime. Some theoretical support for such an approach comes
from the leading order in $1/N_c$ approximation to QCD, which features
just tree diagrams.

    In detail, we calculate the tree diagrams of our model from
an effective chiral Lagrangian which contains resonances but has
interactions with a minimal (for simplicity) number of derivatives.
Hence the initial computed amplitude will be (as in the $1/N_c$
expansion) purely real. This suggests that it is most sensible in the
present approach to compare with the real part of the experimental
amplitude. Of course we still must find a way to ``regularize"
the infinities which arise at the direct channel poles. We interpret
the ``regularization" procedure as equivalent to enforcing unitarity
in the vicinity of the direct channel pole. In the case of a narrow
isolated resonance we adopt the usual Breit--Wigner procedure in
which the offending term in the amplitude is replaced as 

\begin{equation}
\frac{MG}{M^2-s} \rightarrow \frac{MG}{M^2-s-iMG}.
\label{reg1}
\end{equation}
In the case of a very broad resonance we instead replace
\begin{equation}
 \frac{MG}{M^2-s} \rightarrow \frac{MG}{M^2-s-iMG^\prime} ,
\label{reg2}
\end{equation}
where $G^\prime$, which is not required to equal $G$, is taken as
a fitting parameter. Finally, in the case of a ``narrow" resonance
in a non trivial background (characterized by a phase shift $\delta$
in the resonance partial wave) we replace,
\begin{equation}
\frac{MG}{M^2-s} \rightarrow e^{2i\delta}\frac{MG}{M^2-s-iMG} .
\label{reg3}
\end{equation}

    This method can be trivially modified to give a crossing symmetric
invariant amplitude but unitarity may easily be violated in general.
We thus choose the parameters for a putative $\sigma$ meson
represented by (\ref{reg2}) to fit experiment. We then end up with an
amplitude\cite{pipi} which approximately satisfies unitarity and crossing
symmetry.
This is illustrated in a step by step manner in Figs. 1, 2 and 3.
%%%%%%%%%%%%%%%%%%%%%%%%%%%%%%%%%%%%%%%%%%%%%%%%%%%%%%%%%%%%%%%%%%%%%%%%%%%%%%%%%%%%%%%%%%%%%%%%%%%%%%%%%%%%%%%%%%%%%%%%%%%%%%%%%%%%%%%%%%%%%%%%%%%%%%%%%%%%%%%%%%%%%%%%%%%%%%%%%%%%%%%%%%%%%%%%%%%%%%%%%%%
\begin{figure}
\centering
\epsfig{file=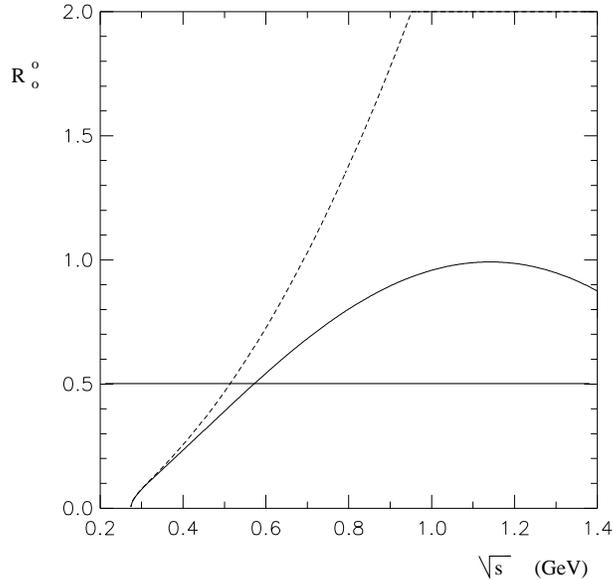,height=3in,angle=0}
\caption{The solid line which shows the current algebra $+ \rho$
result is much closer to the unitary bound of 0.5 than the dashed line
which shows the current algebra result alone.}
\end{figure}

\begin{figure}
\centering
\epsfig{file=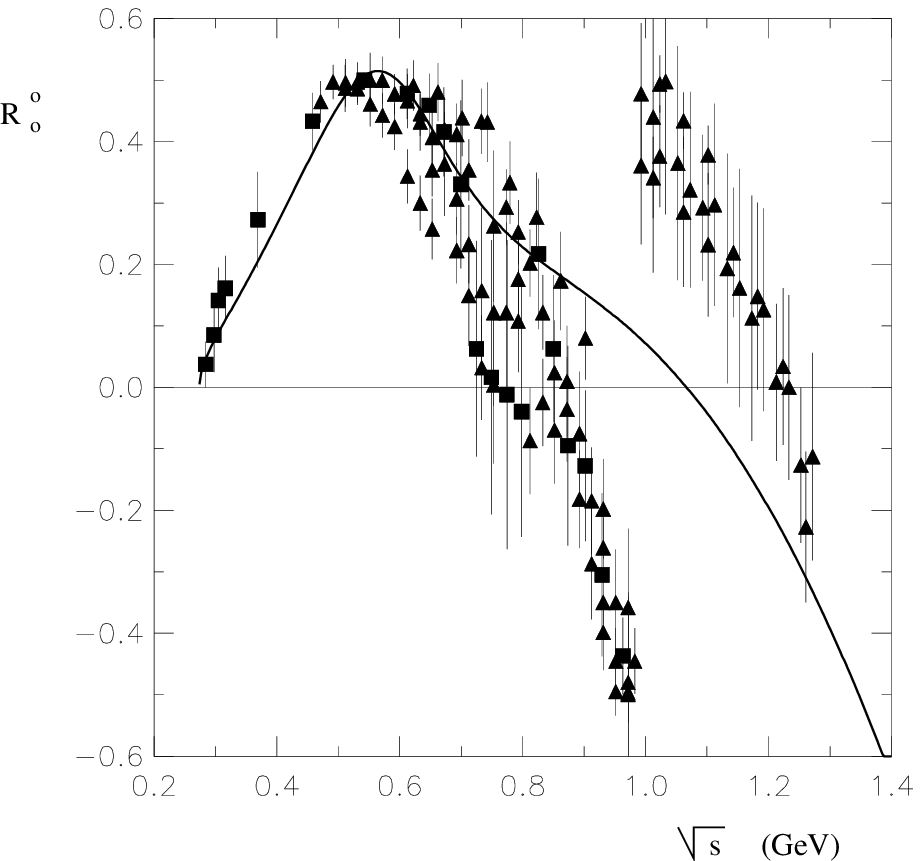,height=3in,angle=0}
\caption{The sum of current algebra $+ \rho + \sigma$ contributions
compared to data.}
\end{figure}

\begin{figure}
\centering
\epsfig{file=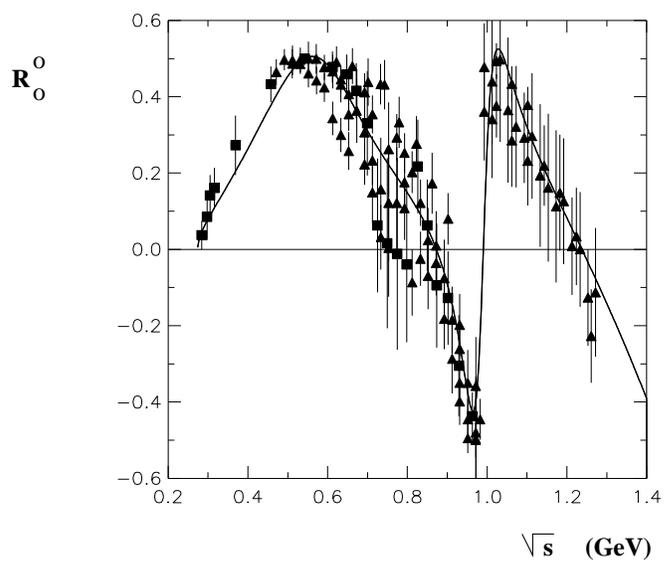,height=3in,angle=0}
\caption{The sum of current algebra $+ \rho + \sigma + f_0(980)$
contributions compared to data.}
\end{figure}

%%%%%%%%%%%%%%%%%%%%%%%%%%%%%%%%%%%%%%%%%%%%%%%%%%%%%%%%%%%%%%%%%%%%%%%%
The real part of the partial wave amplitude $R^0_0$ is obtained by
projecting out the real part of the s--wave $I=0$ component of the 
chiral invariant and crossing symmetric invariant amplitude. We see
in Fig. 1 that the ``current algebra" piece starts violating the
unitarity bound,  $  |R^0_0|  \leq 1/2 $ at about 0.5 GeV and
then runs away. However the inclusion of the  $\rho$
meson exchange diagrams turns the curve in the right direction and
improves, but does not completely cure, the unitarity violation.
This feature, which does not involve any unknown parameters, gives
encouragment to our hope that the cooperative interplay of various pieces
at tree level can explain the low energy scattering. In order to fix up
Fig. 1 we note that the real part of a resonance contribution
vanishes at the pole, is positive before the pole and {\it negative}
above the pole. Thus a scalar resonance with a pole about 0.5 GeV
( where $R^0_0$ in Fig. 1 needs a negative contribution to stay
below $1/2$) should do the job. The result of including such a $\sigma$
meson, parametrized as in (\ref{reg2}), is shown in Fig. 2. Now
note that the predicted $R^0_0(s)$ in Fig. 2 vanishes around 1 GeV.
Thus the phase $\delta$ at 1 GeV (assumed to keep rising) is
about $90^o$ there. Considering this as a background phase for the known
$f_0(980)$, Eq(\ref{reg3}) shows us that the real part of the $f_0(980)$
contribution will get reversed in sign (Ramsauer--Townsend effect).
This is the missing piece in the jig--saw puzzle as Fig. 3 shows. Up
to about 1.2 GeV, the amplitude $R_0^0$ can be explained as the sum of
current algebra, $\rho$ exchange, $\sigma(550)$ exchange and
$f_0(980)$ exchange pieces.

\section{Pi K scattering and $\eta^\prime \rightarrow \eta\pi\pi$}

    A similar treatment was carried out for the $J=0$ partial wave
amplitudes of $\pi K$ scattering\cite{bfss}. In this case the low energy
amplitude is
taken to correspond to the sum of a current algebra contact diagram,
vector $\rho$ and $K^*$ exchange diagrams and scalar $\sigma(550)$,
$f_0(980)$ and $\kappa(900)$ exchange diagrams. The situation in the
interesting $I=1/2$ channel turns out to be very analogous to
the $I=0$ channel of s-wave $\pi\pi$ scattering. Now a $\kappa(900)$
parametrized as in (\ref{reg2}) is required to restore unitarity; it
plays the role of the $\sigma(550)$ in the $\pi\pi$ case. Following
our criterion we expect that to extend this treatment to the 1.5 GeV
region, one should include the many possible exchanges of particles
with masses up to about 1.5 GeV. Nevertheless we found that
a satisfactory description of the 1-1.5 GeV s-wave region is
obtained simply by including the well known $K_0^*(1430)$ scalar
resonance, which plays the role of the $f_0(980)$ in the $\pi\pi$
calculation. 

    The $\eta^\prime \rightarrow \eta\pi\pi$ process is a strong decay
which yields information about the properties of the scalar $a_0(980)$
isovector resonance. The tree diagrams, which are similar to those of
$\pi\eta$ scattering in our model\cite{fs}, include $a_0(980)$,
$\sigma(550)$
and $f_0(980)$ exchanges. Compared to the $\pi\pi$ and $\pi K$ scatterings
there is a simplification in that G-parity invariance prevents vector
meson exchange diagrams from contributing. The associated ``current
algebra" contact diagrams also vanish. It was found that fitting the model
to the experimental Dalitz plot and the rate gave $a_0(980)$
properties consistent with the recent experimental ones.

\section{Scalar nonet ``family" properties}

    The nine states associated with the $\sigma(550)$, $\kappa(900)$,
$f_0(980)$ and $a_0(980)$ are required in order to fit experiment 
in our model. What do their masses and coupling constants suggest
about their quark substructure? (See \cite{putative} for more details.) 
Suppose we first try to assign them
to a conventional $q{\bar q}$ nonet:

\begin{eqnarray}
      \sigma(550)&\sim& \frac{1}{\sqrt{2}}(u{\bar u} + d{\bar
d}),\nonumber\\
      \kappa^+(900)&\sim&u{\bar s},\nonumber\\
       a_0^+(980)&\sim&u{\bar d},\nonumber\\
      f_0(980)&\sim&s{\bar s} .
\label{conven}
\end{eqnarray}
Then there are two puzzles. i) Why aren't the $a_0(980)$ and the
$\sigma(550)$, which have the same number of non--strange quarks,
degenerate? ii) Why aren't these particles, being p--wave states, in the
same 1+ GeV energy region as the other p--wave states?

    To study this, first note that most meson multiplets can be nicely
understood using the concept of ``ideal mixing". In Okubo's
formulation \cite{okubo}, originally applied to the vector meson
multiplet, the meson fields are grouped into a nonet matrix,
\begin{equation}
N_a^b = \left[
\begin{array} {c c c}
N^1_1&a_0^+&\kappa^+\\
a_0^-&N^2_2&\kappa^0\\
{\bar\kappa}^+&{\bar\kappa}^0&N^3_3\\
\end{array}
\right],
\label{nonet}
\end{equation}
where the particle names have been chosen to fit the scalar mesons.
The two $I=0$ states are the SU(3) singlet, $(N^1_1 + N^2_2 +
N^3_3)/\sqrt{3}$ and the SU(3) octet member, $(N^1_1 + N^2_2 -
2N^3_3)/\sqrt{6}$. Okubo's ansatz for the mass terms was,
\begin{equation}
{\cal L}_{mass} = -a{\rm Tr}(NN) - b{\rm Tr}(NN{\cal M}),
\label{mass}
\end{equation}
where $a>0$ and $b$ are real constants and ${\cal M}=diag(1,1,x)$
(with $x=m_s/m_u$) is the ``spurion" matrix which breaks flavor SU(3)
invariance. With (\ref{nonet}) and (\ref{mass}) the $SU(3)$ singlet
and SU(3) octet isoscalar states mix in such a way (ideal mixing)
that the physical mass eigenstates emerge as $(N^1_1 + N^2_2)/\sqrt{2}$
and $N^3_3$. Furthermore there are two mass relations
\begin{eqnarray}
m^2(a_0)&=& m^2(\frac{N^1_1 + N^2_2}{\sqrt{2}}), \nonumber\\
m^2(a_0) - m^2(\kappa) &=& m^2(\kappa) - m^2(N^3_3).
\label{massrelations}
\end{eqnarray}
Note that there are two different solutions depending on the sign of $b$.
If $b>0$ we get Okubo's original case where [with the identifications
$a_0\rightarrow \rho$, $\kappa \rightarrow K^*$, $(N^1_1 + N^2_2)/
\sqrt{2} \rightarrow \omega$ and $N^3_3 \rightarrow \phi$] there
is the conventional ordering
\begin{equation}
m^2(\phi)> m^2(K^*) > m^2(\rho) = m^2(\omega) .
\label{conordering}
\end{equation}
This agrees with counting the number of (heavier) strange quarks
when we identify $N^b_a \sim q_a{\bar q}^b$.

    On the other hand if $b <0$ and we identify $N^3_3 \rightarrow \sigma$
and $(N^1_1 + N^2_2)/\sqrt{2} \rightarrow f_0$, the resulting ordering
would be
 \begin{equation} m^2(f_0) = m^2(a_0) > m^2(\kappa) > m^2(\sigma),
\label{dualordering} \end{equation}
 which is in nice agreement with the
present ``observed" scalar spectrum. But this clearly does not agree with
counting the number of strange quarks while assuming that the scalar
mesons are simple quark anti-quark composites. This unusual ordering will
agree with counting the number of strange quarks if we assume instead that
the scalar mesons are schematically constructed as $N_a^b \sim T_a{\bar
T}^b$ where $ T_a \sim \epsilon_{acd}{\bar q}^c{\bar q}^d$ is a ``dual" 
quark. Specifically
\begin{equation} 
N_a^b \sim T_a{\bar T}^b \sim \left[
\begin{array}{c c c} 
{\bar s}{\bar d}ds&{\bar s}{\bar d}us&{\bar s}{\bar
d}ud \\ 
{\bar s}{\bar u}ds&{\bar s}{\bar u}us&{\bar s}{\bar u}ud \\ 
{\bar
u}{\bar d}ds&{\bar u}{\bar d}us&{\bar u}{\bar d}ud 
\end{array} 
\right]
\label{dualnonet} 
\end{equation}
Note in particular that the light $\sigma
\sim N^3_3$ contains no strange quarks. While this picture seems unusual,
precisely the configuration (\ref{dualnonet}) was found by Jaffe
\cite{jaffe} in the framework of the MIT bag model. The key dynamical
point is that the states in (\ref{dualnonet}) receive (due to the spin and
color spin recoupling coefficients) exceptionally large binding energy 
from the ``hyperfine" piece of the gluon exchange interchange:
\begin{equation}
H_{hf} = - \Delta {\sum}_{i,j}({\bf S}_i\cdot{\bf S}_j)({\bf
F}_i\cdot{\bf
F}_j),
\label{hyperfine}
\end{equation}
wherein the sum goes over all pairs $i,j$ while $S_i$ and $F_i$ are
respectively the spin and color generators acting on the $i^{th}$
quark or antiquark.

    While the picture above seems close to our expectations it is not
quite right in detail. For example the masses do not exactly obey 
(\ref{massrelations}). Furthermore the simplest model for decay would give
that $f_0 \rightarrow \pi\pi$ vanishes, in contradiction to experiment.
Hence we add the extra mass terms
\begin{equation}
{\cal L}_{mass} = {\rm Eq.}(\ref{mass}) - c{\rm Tr}(N)Tr(N)
-d{\rm Tr}(N){\rm Tr}(N{\cal M}).
\label{fullmass}
\end{equation}
The $c$ and $d$ terms give $f_0-\sigma$ mixing. Now we solve for
$(a,b,c,d)$ in terms of the four masses $m_\sigma=$550 MeV,
$m_\kappa=$900MeV, $m_{a_0}=$983.5 MeV and $m_{f_0}=$980 MeV. The
solution boils down to a quadratic equation for (say) $d$. This gives two
possible values for the mixing angle $\theta_s$ defined by,
\begin{equation}
\left( \begin{array}{c} \sigma\\ f_0 \end{array} \right) = \left(
\begin{array}{c c} {\rm cos} \theta_s & -{\rm sin} \theta_s \\ {\rm sin}
\theta_s & {\rm cos} \theta_s \end{array} \right) \left( \begin{array}{c}
N_3^3 \\ \frac {N_1^1 + N_2^2}{\sqrt 2} \end{array} \right).
\label{sf_mix}
\end{equation}
The solution $\theta_s \approx -90^o$, giving $\sigma \approx (N^1_1
+ N^2_2)/\sqrt{2}$ seems to correspond to restoring the $q{\bar q}$ model
(\ref{conven}) for the scalars once more. The other solution $\theta_s
\approx -20^o$ corresponds to $\sigma$ being mainly $N^3_3$ which was just
noted to be a characteristic signature of the $qq{\bar q}{\bar q}$
model (\ref{dualnonet}). The very existence of these two different 
solutions highlights the fact that by just assuming a flavor
transformation property for the scalars we are not forcing a
particular identification of their underlying quark structure.
Different substructures are naturally associated with different
values of the parameters in the same effective Lagrangian. In
any event, the extra terms in (\ref{fullmass}) have restored the ambiguity
about the scalars' structure. We need more information to decide the
issue. For this purpose we look at the trilinear couplings.

    Using SU(3) invariance we write
\begin{eqnarray}    
{\cal L}_{N\phi \phi} =
&A&{\epsilon}^{abc}{\epsilon}_{def}N_{a}^{d}{\partial_\mu}{\phi}_{b}^{e}
{\partial_\mu}{\phi}_{c}^{f}
+ B {\rm Tr} \left( N \right) {\rm Tr} \left({\partial_\mu}\phi
{\partial_\mu}\phi \right)  \nonumber \\ 
&+&C {\rm Tr} \left( N {\partial_\mu}\phi
\right) {\rm Tr} \left( {\partial_\mu}\phi \right) 
+ D {\rm Tr} \left( N \right) {\rm Tr}
\left({\partial_\mu}\phi \right)  {\rm Tr} \left( {\partial_\mu}\phi
\right),
\label{trilinear}
\end{eqnarray}
where $A,B,C,D$ are four real constants and $\phi$ represents the usual
pseudoscalar nonet matrix. 
The derivatives stem from the 
requirement that (\ref{trilinear}) be the leading part of a chiral
invariant object. If desired, we can rewrite the $A$ term as a linear
combination of the usual Tr$(N\partial_\mu\phi\partial_\mu\phi)$
and the three other terms. The motivation for the form given is
that by itself the $A$ term yields zero for $f_0 \rightarrow \pi\pi$
and $\sigma \rightarrow K{\bar K}$, both of which should vanish in the
``fall apart" picture of a $T{\bar T}$ type scalar meson. Note that all
the
coupling constants which enter into our treatment of $\pi\pi$ and 
$\pi K$ scattering depend on just $A$ and $B$; $C$ and $D$ contribute only
to the decays containing $\eta$ or $\eta^\prime$ in the final state.
For examples of couplings:
\begin{eqnarray}
\gamma_{\kappa K\pi}&=&\gamma_{a_0KK}=-2A, \nonumber \\
\gamma_{\sigma\pi\pi}&=& 2B{\rm sin}
\theta_s-\sqrt{2}(B-A){\rm cos}\theta_s, etc.
\label{examples}
\end{eqnarray}
 
    The mixing angle solution which best fits the couplings needed to
explain the $\pi\pi$ and $\pi K$ scattering turns out to be $\theta_s
\approx -20^o$. Together with a suitable choice of $C$ and $D$, the
interactions involving $\eta$ and $\eta^\prime$ are also consistently
described (as mentioned in section III). Thus it seems that our results
point to a picture in which the light scalars are mainly dual quark-
dual antiquark rather than quark-antiquark type. Very recently
Achasov \cite{achasov} has argued that new experimental data from
Novosibirsk
on the radiative decay $\phi(1020) \rightarrow \pi^0\eta\gamma$
are better fit with a $qq{\bar q}{\bar q}$ type model of the $a_0(980)$.

To sum up: assuming $\sigma(550)$, $\kappa(900)$, $f_0(980)$ and
$a_0(980)$ to belong to a nonet $N_a^b$ which is fitted into a chiral
Lagrangian, we have found the parameters $A,B,C,D$ which specify sixteen
scalar-pseudoscalar-pseudoscalar coupling constants. These couplings and
masses are used to explain $\pi\pi$ scattering ($\sigma, f_0$),
$\pi K$ scattering ($\kappa$) and $\eta^\prime \rightarrow \eta\pi\pi$
($a_0$) with regularized tree amplitudes. Furthermore, a small
$\sigma-f_0$ mixing angle in (\ref{sf_mix}) suggests that $N_a^b$
is describing a structure similar to a dual quark-dual antiquark. If
this picture is correct there are many interesting applications
and questions.

\section{Possible mechanism for next lowest-lying scalars}

Of course, the success of the phenomenological quark model suggests
that there exists a nonet of ``conventional" $q{\bar q}$ scalars in
the 1+ GeV range. What are the experimental candidates for these?
\cite{rpp} The situation for the isoscalar candidates is presently
confusing. The $f_0(1370)$ may actually correspond to two different
states. The $f_0(1500)$ may be a glueball while the $f_J(1710)$ does
not necessarily have spin zero. Thus we will not focus on the
isoscalars now. On the other hand the Review of Particle Physics
``endorses" the isovector and isospinor candidates
\begin{eqnarray}
&a_0(1450)& : M=1474 \pm 19 {\rm MeV},\hskip.5cm \Gamma=265 \pm 13
{\rm MeV},
\nonumber \\
&K_0^*(1450)& : M=1429 \pm 6 {\rm MeV},\hskip.5cm \Gamma=287 \pm 23
{\rm MeV}.
\nonumber
\end{eqnarray}

On the way to taking these states seriously as members of an ordinary
p-wave nonet we encounter three puzzles. i) The mass of the $a_0^+(1450)$
(presumably a $u{\bar d}$ state is greater than that of the
$K_0^{*+}(1430)$ (presumably a $u{\bar s}$ state). ii) The $a_0(1450)$
and $K_0^*(1430)$ are not less massive than the corresponding p-wave
tensor mesons $a_2(1320)$ and $K_2^*(1430)$, as expected from an
$L\cdot S$ interaction (e.g. $m[\chi_{c2}(1p)]>m[\chi_{c0}(1p)]$).
iii) Assuming the known decay modes $K_0^*(1430) \rightarrow K\pi$
and $a_0(1450) \rightarrow \pi\eta, K{\bar K}, \pi\eta^\prime$
saturate the total widths, we have from SU(3) flavor invariance
that $\Gamma[a_0(1450)] = 1.51 \Gamma[K_0^*(1430)]$. However,
experimentally it is $(0.92 \pm 0.12)\Gamma[K_0(1430)]$ instead.

These puzzles can be simply resolved\cite{bfs} if we assume that 
an ideally mixed heavier $q{\bar q}$ nonet $N^\prime$ in turn mixes with
an ideally mixed $T{\bar T}$ nonet $N$ (as in (\ref{dualnonet})) via
\begin{equation}
{\cal L}^\prime = - \gamma {\rm Tr}(NN^\prime).
\label{nonetmixing}
\end{equation}
Actually the assumption of exact ideal mixings is a simplification which
can be relaxed. The mechanism is driven by the fact that $m(a_0^\prime)
<m(K_0^\prime)$ while $m(a_0)>m(K_0)$. Here the subscript zero refers
to the unmixed $N$ and $N^\prime$ members. The splittings are summarized
in Fig. 4.

%%%%%%%%%%%%%%%%%%%%%%%%%%%%%%%%%%%%%%%%%%%%%%%%%%%%%%%%%%%%%%%%%%%%%%%%%
\begin{figure}
\centering
\epsfig{file=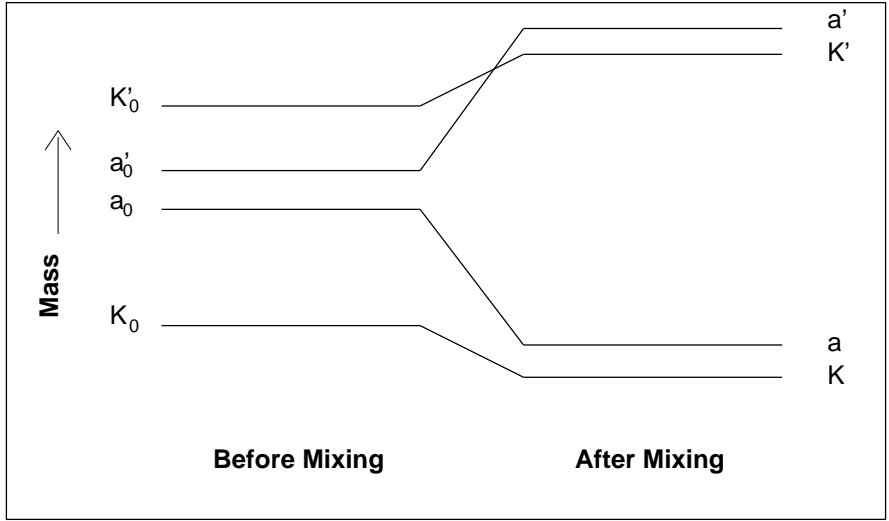,height=3in,angle=0}
\vskip 0.3cm
\caption{Mixing of two nonets-a',K',a and K stand respectively for the
"physical" states $a_0(1450), K_0^*(1430), a_0(980)$ and $\kappa(900)$.
$K_0$ and $a_0$ are the unmixed isospinor and isovector $qq{\bar q}
{\bar q}$ states, while $K_0^\prime$ and $a_0^\prime$ are the
corresponding unmixed $q{\bar q}$ states.}
\end{figure}

%%%%%%%%%%%%%%%%%%%%%%%%%%%%%%%%%%%%%%%%%%%%%%%%%%%%%%%%%%%%%%%%%%%%%%%%%
The explanations are: i)Think of a perturbation theory approach. There is
a smaller ``energy denominator" for $a_0-a_0^\prime$ mixing than for 
$K_0-K_0^\prime$ mixing. Thus there is more $a_0-a_0^\prime$ repulsion as
shown in Fig. 4. ii) Since the mixing of two levels ``repels" them,
both $a_0(1450)$ and $K_0^*(1430)$ are heavier than would be expected
otherwise. Similarly the light scalars $a_0(980)$ and $\kappa(900)$
are lighter than they would be without the mixing (\ref{nonetmixing}).
iii) The difference between the $a_0(1450)$ and $K_0^*(1430)$
decay coupling constants can be understood from the necessarily greater
mixture of the $qq{\bar q}{\bar q}$ component in the $a_0(1450)$
than in the $K_0^*(1430)$.

Clearly, looking at the isoscalars will be especially interesting when the
experimental situation becomes clearer.

   We would like to thank Francesco Sannino and Masayasu Harada
for fruitful collaboration. One of us (J.S.)
would like to thank the organizers for arranging a stimulating and
enjoyable conference. The work has been supported in part by the
US DOE under contract DE-FG-02-85ER40231.

\end{document}